\documentclass[aps,prc,twocolumn,superscriptaddress]{revtex4-1}
\usepackage{graphicx}
\usepackage{amsmath}
\usepackage{braket}
\usepackage{url}
\usepackage{multirow}
\usepackage{multirow}
\usepackage{amssymb}
\usepackage{booktabs}
\usepackage{rotating}
\usepackage{bm}
\usepackage{bbm}
\usepackage{ulem}

\usepackage[colorlinks,
linkcolor=blue,
anchorcolor=red,
urlcolor=red,
citecolor=red]{hyperref}

\begin{document}

\title{Investigation of isospin-symmetry-breaking in mirror energy difference and nuclear mass with \textit{ab initio} calculations}

\author{H. H. Li}%\thanks{These authors equally contributed to this work.}
\affiliation{CAS Key Laboratory of High Precision Nuclear Spectroscopy, Institute of Modern Physics,
Chinese Academy of Sciences, Lanzhou 730000, China}
\affiliation{School of Nuclear Science and Technology, University of Chinese Academy of Sciences, Beijing 100049, China} 
\author{Q. Yuan} %\thanks{These authors equally contributed to this work.}
\affiliation{School of Physics,  and   State Key  Laboratory  of  Nuclear  Physics   and  Technology, Peking University, Beijing  100871, China}
\author{J. G. Li}\email[]{jianguo\_li@impcas.ac.cn}
\affiliation{CAS Key Laboratory of High Precision Nuclear Spectroscopy, Institute of Modern Physics,
Chinese Academy of Sciences, Lanzhou 730000, China}
\affiliation{School of Nuclear Science and Technology, University of Chinese Academy of Sciences, Beijing 100049, China}
\author{M. R. Xie}
\affiliation{CAS Key Laboratory of High Precision Nuclear Spectroscopy, Institute of Modern Physics,
Chinese Academy of Sciences, Lanzhou 730000, China}
\author{S. Zhang}
\affiliation{School of Physics,  and   State Key  Laboratory  of  Nuclear  Physics   and  Technology, Peking University, Beijing  100871, China}
\author{Y. H. Zhang}
\affiliation{CAS Key Laboratory of High Precision Nuclear Spectroscopy, Institute of Modern Physics,
Chinese Academy of Sciences, Lanzhou 730000, China}
\affiliation{School of Nuclear Science and Technology, University of Chinese Academy of Sciences, Beijing 100049, China} 
\author{X. X. Xu}
\affiliation{CAS Key Laboratory of High Precision Nuclear Spectroscopy, Institute of Modern Physics,
Chinese Academy of Sciences, Lanzhou 730000, China}
\affiliation{School of Nuclear Science and Technology, University of Chinese Academy of Sciences, Beijing 100049, China} 
\author{N. Michel}
\affiliation{CAS Key Laboratory of High Precision Nuclear Spectroscopy, Institute of Modern Physics,
Chinese Academy of Sciences, Lanzhou 730000, China}
\affiliation{School of Nuclear Science and Technology, University of Chinese Academy of Sciences, Beijing 100049, China}
\author{F. R. Xu}
\affiliation{School of Physics,  and   State Key  Laboratory  of  Nuclear  Physics   and  Technology, Peking University, Beijing  100871, China}
\author{W. Zuo}
\affiliation{CAS Key Laboratory of High Precision Nuclear Spectroscopy, Institute of Modern Physics,
Chinese Academy of Sciences, Lanzhou 730000, China}
\affiliation{School of Nuclear Science and Technology, University of Chinese Academy of Sciences, Beijing 100049, China}
%\affiliation{Institute of Modern Physics, Chinese Academy of Sciences, Lanzhou 730000, China}
%\affiliation{School of Nuclear Science and Technology, University of Chinese Academy of Sciences, Beijing 100049, China}
%\email[]{nicolas.michel@impcas.ac.cn}
\date{\today}

\begin{abstract}

Isospin-symmetry breaking is responsible for the energy difference of excited states in mirror nuclei. It also influences the coefficient of the isobaric multiplet mass equation. In the present work, we extensively investigate isospin-symmetry breaking in medium mass nuclei within \textit{ab initio} frameworks. For this, we employ the \textit{ab initio} valence-space in-medium similarity renormalization group  approach, in which charge-symmetry and charge-independence breakings are included in the adopted  nuclear force.
The mirror energies of $sd$- and $pf$-shell nuclei are computed for that matter.
The effects of single-particle states on weakly bound and unbound nuclear states, especially those of the $s$-wave, are discussed. Predictions are also made concerning proton drip-line nuclei bearing large mirror energy difference.
Finally, the coefficient of the isobaric multiplet mass equation in $T=1/2$ and $T=1$ isospin multiplets for $A=18$ to $A=76$ is calculated.  

%make predictions about unknown sd- and fp-shell nuclei of experimental interest. 

%Our calculations overall satisfy the existing experiment which indicates the importance of three-body force and denotes conclusions. The occupation of $1\textit{s}_{1/2}$ + $0\textit{d}_{3/2}$ orbits particularly $1\textit{s}_{1/2}$ orbit greatly influence the mirror energy difference. The $J$ dependence of mirror energy difference and triplet energy difference don't vary monotonously as the spin. Coulomb force, charge symmetry breaking and charge independent breaking are three main factors resulting in isospin-symmetry breaking from isobaric multiplet mass equation. 

\end{abstract}

%\keywords{\textit{Ab initio} Isospin symmetry breaking}

\maketitle

\section{Introduction}
%\textit{Introduction.}~
Isospin symmetry of nuclear forces is a fundamental assumption in nuclear physics, where it has been seen to be almost exact \cite{BENTLEY2007497,Heisenberg1932berDB, PhysRev.51.106} . It leads to the introduction of isospin by Heisenberg, as the proton and neutron can be viewed as two different states of the same particle,  differing only by the projection of isospin ($t_z$) \cite{Heisenberg1932berDB, PhysRev.51.106}. Under this assumption and irrespective of electromagnetic effects, i.e.~with exact isospin symmetry, mirror nuclei, whose protons and neutrons are interchanged, should bear the same energy levels. 

The differences in excitation energy of analogue states in mirror nuclei follow this rule up to a few tens or hundreds of keVs \cite{PhysRevLett.89.142502,2006ARNPS..56..253M}. In particular, there is a large mirror energy difference (MED) in the mirror nuclei states where the valence protons of proton-rich nuclei occupy weakly-bound or unbound $s$ or $p$-wave, which is called the Thomas-Ehrman shift (TES) \cite{PhysRev.88.1109, PhysRev.81.412}.
Indeed, with the developments of accelerator and detector, measurements of excited states of drip-line nuclei become possible \cite{PhysRevLett.92.132502,PhysRevLett.97.132501,PhysRevLett.97.152501,PhysRevC.82.061301,PhysRevLett.117.082502,FERNANDEZ2021136784}. A typical example is the $3_1^+$ state in $^{18}$Ne and $^{18}$O mirror nuclei \cite{PhysRevLett.83.45}.
Another is the mass difference of isobaric multiplets which cannot be explained from Coulomb effects alone, and is called the Nolen-Schiffer anomaly (NSA) \cite{doi:10.1146/annurev.ns.19.120169.002351}. While Coulomb effects are  the main reason for isospin-symmetry breaking (ISB), theoretical investigations indicated that the charge dependence of nuclear forces also plays a role therein.
Experimental nucleon-nucleon phase shifts have shown that the neutron-neutron interaction ($V_{nn}$) is about 1\% larger than the proton-proton interaction ($V_{pp}$), and that the proton-neutron interaction ($V_{pn}$) is about 2.5 \% stronger than the average of $V_{nn}$ and $V_{pp}$ interactions \cite{MACHLEIDT20111}. These are denoted as the charge-symmetry breaking (CSB) and charge-independence breaking (CIB), respectively \cite{MACHLEIDT20111}. In fact, the large MED and isobaric multiplet mass equation (IMME) can be treated as probes to investigate ISB.
So far, two theoretical methods have been applied to explain how ISB arises, i.e., standard shell model \cite{BENTLEY2007497,PhysRevLett.89.142502, PhysRevLett.110.172505, PhysRevC.87.054304} and density functional theory \cite{BACZYK2018178,PhysRevLett.124.152501}. Parameters are introduced and constrained by data in those two models \cite{BENTLEY2007497,BACZYK2018178,PhysRevC.106.024327}.

Over the past decade, the development of \textit{ab initio} frameworks exhibited great progress in nuclear physics thanks to the introduction of chiral interactions issued from effective field theory \cite{HERGERT2016165,doi:10.1146/annurev-nucl-101917-021120,BARRETT2013131,RevModPhys.87.1067,Hagen_2014,PhysRevLett.105.032501,PhysRevLett.110.242501,Drischler2021ABA}. In particular, the CSB and CIB effects can be well treated with nuclear interactions devised from effective field theory (see details in Ref. \cite{MACHLEIDT20111}). 
Of particular interest is the \textit{ab initio} valence-space in-medium similarity renormalization group (VS-IMSRG) method \cite{PhysRevLett.118.032502,PhysRevC.85.061304,PhysRevLett.106.222502},
which can build effective nuclear many-body Hamiltonians in relatively small valence spaces via a continuous unitary transformation \cite{PhysRevC.85.061304,HERGERT2016165,doi:10.1146/annurev-nucl-101917-021120,WEGNER200177,PhysRevC.75.061001,PhysRevC.87.021303,BOGNER201094,PhysRevC.105.034333}.  
Both closed- and open-shell nuclei can be implemented in the framework of VS-IMSRG.
Thus, this technology allows to investigate ISB in the ground and excited states of many nuclei.

%The ISB has extensive applications to nuclei, such as in mapping the proton drip line \cite{PhysRevLett.110.172505, PhysRevC.55.2407,BLANK2008403}, determining the $rp$- process path \cite{PhysRevC.87.065803, PhysRevC.95.055806}, testing the standard model \cite{PhysRevLett.83.1299, smirnova:hal-02266319,PhysRevLett.94.092502,PhysRevLett.97.102501,PhysRevLett.112.102502,particles4040038}. We firstly investigated the MEDs in sd-shell to explain ISB. There is a great coincidence between our results and experiments. Results give that the large MEDs come from the occupation of $1\textit{s}_{1/2}$ orbit. We also calculated MED and triplet energy differences (TED) for mass number $A=46$ with $T=1$ triplets, giving that variations of MED and TED with spin don't be monotonous. Then we predicted spectrum and MEDs of experimentally unmeasured concerned  $^{21}$Al, $^{22}$Si, $^{23}$Si, $^{27}$P nuclei. Finally, we researched isobaric analog states (IAS) for $T=1/2$ doublets and $T=1$ triplets, calculated coefficients of IMME. These results have a good agreement with existing experiments and can provide credible mass predictions for unknown mass nuclei, especially for neutron-deficient nuclei which near the proton drip line.
 
This paper is structured as follows. Firstly, the \textit{ab initio} VS-IMSRG method is briefly introduced. Then, the  MED in $sd$- and $pf$-shell nuclei are investigated, emphasizing the TES in $sd$-shell many-body states. Afterwards, the calculated coefficients for IMMEs of $T=1/2$ isospin doublets and $T=1$ isospin triplets $sd$- and $pf$-shell nuclei are presented. Then, one proceeds to the summary of the paper.

%  We initially accounted for SRG to IMSRG and finally VS-IMSRG evolution. In overall consideration of computerized limitations and theoretical reliability, we took VS-IMSRG(2) approximation which normal-ordering two-body machinery contains contributions of three-body force. In the part of "\textit{Results}", conclusions were made about the occupation of large MEDs, $J$ dependence of MED and TED as well as the coefficients of IMME in $T=1/2$ and $T=1$ comparing current experimental situations, and we analyzed the causes of isospin-symmetry breaking in theory. We also predicted the characters of unknown sd-shell nuclei experimentalists concerning. The final part is the section "$Summary$".

\section{Method}
The intrinsic $A$-nucleon Hamiltonian reads :
\begin{equation}
 H=\sum_{i=1}^{A}\left(1-\frac{1}{A}\right) \frac{\boldsymbol{p}_{i}^{2}}{2 m}+\sum_{i<j}^{A}\left(v_{i j}^{\mathrm{NN}}-\frac{\boldsymbol{p}_{i} \cdot \boldsymbol{p}_{j}}{m A}\right)+\sum_{i<j<k}^{A} v_{i j k}^{3 \mathrm{N}},
 \label{H_intrinsic}
\end{equation}
where $\boldsymbol{p_i}$ is the nucleon momentum in the laboratory, and $m$ is mass of the nucleon. 
$v^{\rm NN}$ and  $v^{\rm 3N}$ are the two-body (NN) and three-nucleon (3N) interactions, respectively.
The well established NN+3N interaction provided by the 1.8/2.0 (EM) potential is used in our present work, as it has been shown to systematically reproduce ground-state energies up to $^{132}$Sn \cite{PhysRevLett.126.022501,PhysRevLett.120.152503,PhysRevC.105.014302}. For the 1.8/2.0 (EM) potential, the initial chiral next-to-next-to-next-to-leading order (N$^3$LO) NN force \cite{PhysRevC.68.041001} is softened by a similarity renormalization group (SRG) evolution \cite{PhysRevC.75.061001} using $\lambda_{\rm SRG} = 1.8$ fm$^{-1}$, where a cutoff $\Lambda = 2.0$ fm$^{-1}$ is chosen for the corresponding next-to-next-to-leading order
(N$^2$LO) 3N interaction. The short-range low-energy constants $c_D$ and $c_E$ are optimized to reproduce the triton bounding energy and $^{4}$He radius \cite{PhysRevC.83.031301}.
Within the chiral EFT framework, charge-symmetry and charge-independence breaking effects are considered among others via $\pi$-mass splitting in the pion-exchange, pion-nucleon coupling constant, nucleon-mass splitting, electromagnetic corrections, low-energy constants of contact terms, and so on. A more comprehensive and systematic study can be found in Ref. \cite{PhysRevC.72.044001,MACHLEIDT20111}. The Coulomb force is also included in Eq. (\ref{H_intrinsic}). In practical calculations, the harmonic-oscillator (HO) basis is used for the definitions of the model space. We consider $\hbar \omega =16$ MeV and 15 HO major shells are included (i.e.~$e=2n+l \leq e_{\mathrm{max}}=14$), and the  HO energies in the three-body sector are limited as well to $e_{3\text{max}} = 2n_a+2n_b+2n_c+l_a+l_b+l_c\leq 14$.

It is convenient to rewrite the Hamiltonian of Eq. (\ref{H_intrinsic}) with normal ordering with respect to the single determinant or ensemble
reference state $|\Phi\rangle$ \cite{PhysRevLett.118.032502}:
\begin{equation}
 \begin{aligned}
 H= & E+\sum_{ij} f_{ij}: a_{i}^{\dagger} a_{j}:+\frac{1}{4} \sum_{ijkl} \Gamma_{ijkl}: a_{i}^{\dagger} a_{j}^{\dagger} a_l a_k :\\
  &+\frac{1}{36} \sum_{i j k l m n} W_{i j k l m n}: a_{i}^{\dagger} a_{j}^{\dagger} a_{k}^{\dagger} a_{n} a_{m} a_{l}:,
 \label{H_T2}
  \end{aligned}
\end{equation}
where the strings of creation and annihilation operators obey $\langle\Phi|: a_{i}^{\dagger} \cdots a_{j}:| \Phi\rangle=0$. Indeed, the normal-ordered zero-, one-, and two-body parts, i.e.~$E$, $f_{ij}$ and $\Gamma_{ijkl}$, contain the main contributions of $v^{3N}$, so that one can neglect the numerically expensive normal-ordered three-body part $W_{ijklmn}$ of the Hamiltonian  \cite{HERGERT2016165,doi:10.1146/annurev-nucl-101917-021120}. 

In its initial studies, IMSRG had been applied to calculate doubly magic nuclei \cite{PhysRevLett.106.222502}. In order to calculate open-shell nuclei, it is necessary to divide the single-particle Hilbert space into core, valence, and excluded spaces. This doing, the construction of an effective valence-space Hamiltonian can be performed, so that its diagonalization in the considered model space can provide an {\it ab initio} description for nuclear structure. VS-IMSRG method aims at constructing such a valence-space effective Hamiltonian \cite{PhysRevC.85.061304, PhysRevLett.118.032502}. Calculations of nuclear observables and investigations of ISB are then made possible. 
%To this end, the “off-diagonal” parts of the Hamiltonian is defined as $H^{od}=\{f_{ph},f_{qv},f_{pp'},f_{hh'}, \Gamma_{pp'hh'},\Gamma_{pp'(vh\ {\rm or}\ hv)},\Gamma_{(pq \ or \ qp)vv'}\}$ plus Hermitian conjugates \cite{PhysRevC.85.061304}, where indices $h$, $v$, and $q$ indicate core, valence, and excluded-space
%orbits, respectively, and $p$ indicates either $v$ or $q$. 
The decoupling can be achieved by solving the flow equation
\begin{equation}
\frac{dH(s)}{ds}=[\eta(s),H(s)],
\label{FE}
\end{equation}
with the anti-Hermitian generator 
\begin{equation}
\eta(s)\equiv\frac{dU(s)}{ds}U^{\dagger}(s)=-\eta^{\dagger}(s),
\end{equation}
where $U(s)$ is the unitary transformation operator.

In this work, we will use VS-IMSRG with ensemble normal-ordering (ENO) \cite{PhysRevLett.118.032502,imsrg_code} to generate the valence-space Hamiltonian, whereby the VS-IMSRG code of Ref.\cite{imsrg_code} is utilized for that matter. The obtained Hamiltonian can then be exactly diagonalized using the shell model code of Ref.~\cite{MICHEL2020106978}.  In the present work, the $A=18$ to $A=75$ isotopes are calculated, using full $sd$- and $pf$-shell valence spaces for both valence protons and valence neutrons, respectively.

%\clearpage
\section{Results}
\subsection{Mirror Energy Differences}
%\textit{Results.}~
%Mirror nuclei should have the same level with isospin symmetry approximation.
%However, due to the isospin non-conserving and Coulomb forces, the symmetry is broken. 
Isospin-symmetry breaking implies MED, i.e.~the excited state energy difference between analogue states in mirror nuclei with the same mass number $A$, total isospin $T$ and spin-parity $J^{\pi}$ but different isospin third component $T_z$, is not equal to zero  \cite{PhysRevLett.89.142502,PhysRevLett.92.132502,PhysRevLett.97.132501,PhysRevLett.97.152501,PhysRevLett.110.172505,KANEKO2017521}. 
TES also occurs at proton dripline, where proton-rich isotopes exhibit a large MED. This provides a powerful probe to access the origin of ISB \cite{PhysRevLett.89.142502,PhysRevLett.92.132502,PhysRevLett.97.132501,PhysRevLett.97.152501,PhysRevLett.110.172505,KANEKO2017521} and further information about nuclear structure \cite{BENTLEY2007497,PhysRevLett.121.032502}. The MED is given by 
%Mirror nuclei should have the same level schemes due to the isospin symmetry, the mirror energy difference (MED), i.e., the excited state energy difference between mirror nuclei with the same mass number $A$, total isospin $T$ and spin-parity $J^{\pi}$ but different isospin third component $T_z$, becomes thus a powerful probe to access the isospin-symmetry breaking} \cite{PhysRevLett.89.142502,PhysRevLett.92.132502,PhysRevLett.97.132501,PhysRevLett.97.152501,PhysRevLett.110.172505,KANEKO2017521} and further the information of nuclear structure \cite{BENTLEY2007497,PhysRevLett.121.032502}. The MED is given by
\begin{equation}
\text{MED}(A, T)=E_{ex}\left(T, T_{z_>}\right)-E_{ex}\left(T, T_{z_<}\right),
 \label{ME}
\end{equation}
in which the $E_{ex}$ is the excitation energy and $T_{z_>}$ ($T_{z_<}$) refers to the nucleus of largest (smallest) isospin projection considered in the MED.

The TES is principally caused by the coupling to weakly bound or unbound $s$- and $p$- orbitals. Indeed, the absent or small centrifugal parts implies that the proton and neutron states close to particle-emission threshold have extended wave functions in coordinate space. Thus, mirror nuclear states whose $s$ and $p$ components are important have different wave function asymptotes, which is responsible for the  generating TES. 
%The unbound $0p_{3/2}$ and $0p_{1/2}$ single-particle orbits provide visible TES states in $p$-shell nuclei, such as in $^{11}$O/$^{11}$Li \cite{PhysRevLett.122.122501} and $^{13}$C/$^{13}$N \cite{ensdf}.
An inversion of ground states occurs in the $^{16}$F and $^{16}$N mirror nuclei, which is mainly due to the unbound proton $1s_{1/2}$ orbital \cite{PhysRevC.90.014307,PhysRevC.106.L011301}. Rich information related to TES has been obtained in the $sd$-shell proton drip-line nuclei, where many states with large TES have been observed. One can give the examples of the mirror pairs $^{18}$Ne/$^{18}$O \cite{PhysRevLett.102.152502,ZHANG2022136958} and $^{22}$Al/$^{22}$F \cite{PhysRevLett.125.192503}.
In the $sd$-shell nuclei, TES is driven by $s$-waves. Indeed, the proton $1s_{1/2}$ orbital is weakly bound or unbound in proton drip-line nuclei, whereas the neutron $1s_{1/2}$ is well-bound in their mirror neutron-rich nuclei. However, few calculations have been done due to the difficulty to include both Coulomb and isospin-non-conserving forces in theoretical models. We could circumvent this problem by employing the  \textit{ab-initio} VS-IMSRG, in which the Coulomb and isospin-non-conserving forces are exactly taken into account. We can then investigate the large MED states occurring in $sd$-shell nuclei.

\begin{figure}[!htb]
\includegraphics[width=1.00\columnwidth]{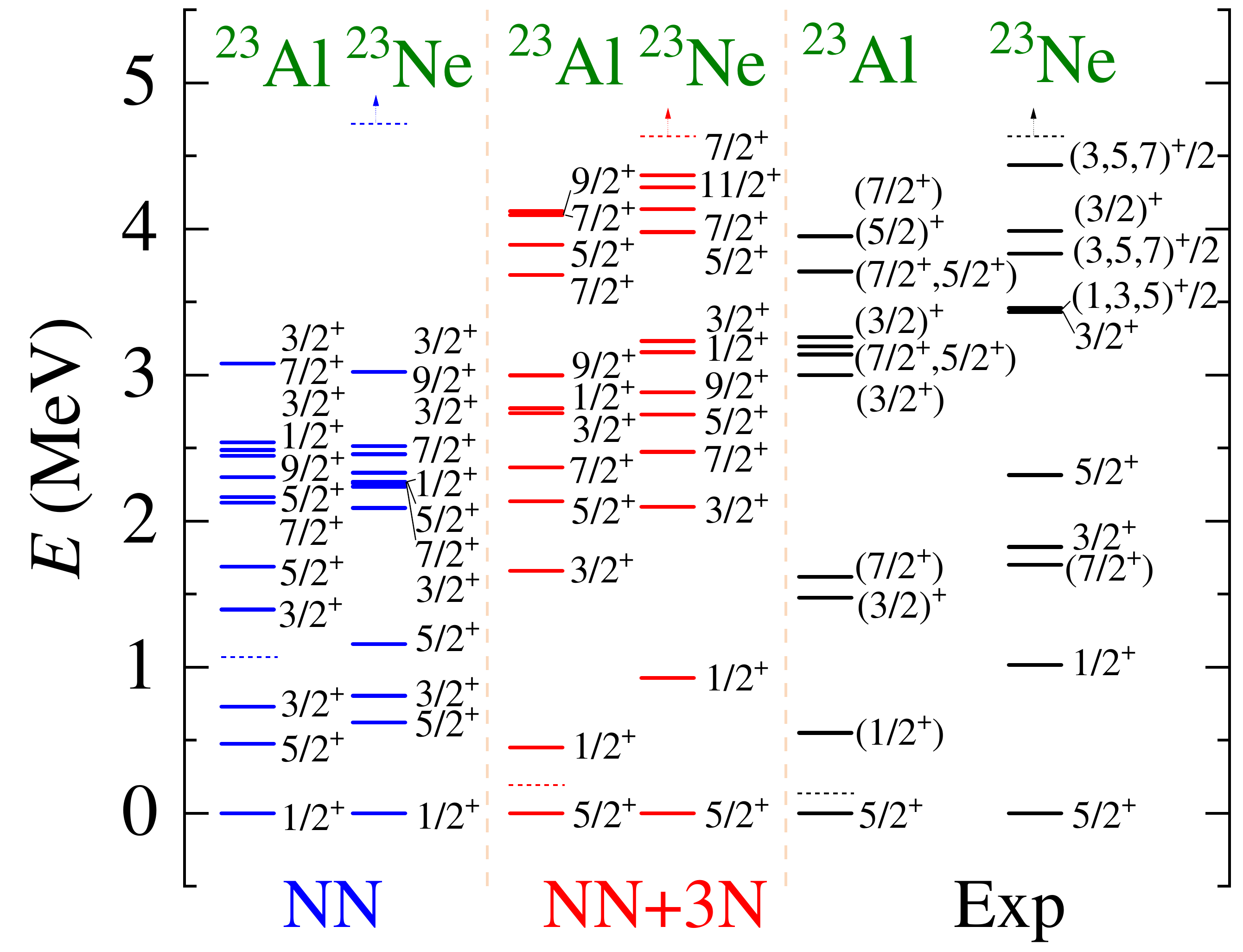}% Here is how to import EPS art
\caption{\textit{Ab initio} VS-IMSRG calculations of spectra of mirror nuclei $^{23}$Al and $^{23}$Ne with only NN and NN + 3N interactions, along with experimental data \cite{ensdf}. The calculations with 1.8/2.0 (EM) two- and three-nucleon potential and only SRG $\lambda =$ 1.8 fm$^{-1}$ two-nucleon potential are labeled by NN + 3N and NN, respectively. The one-proton and neutron separation energies of $^{23}$Al and $^{23}$Ne, respectively, are indicated by the dashed lines with arrows.}
\label{23Al}
\end{figure}

\begin{figure}[!htb]
\includegraphics[width=1.00\columnwidth]{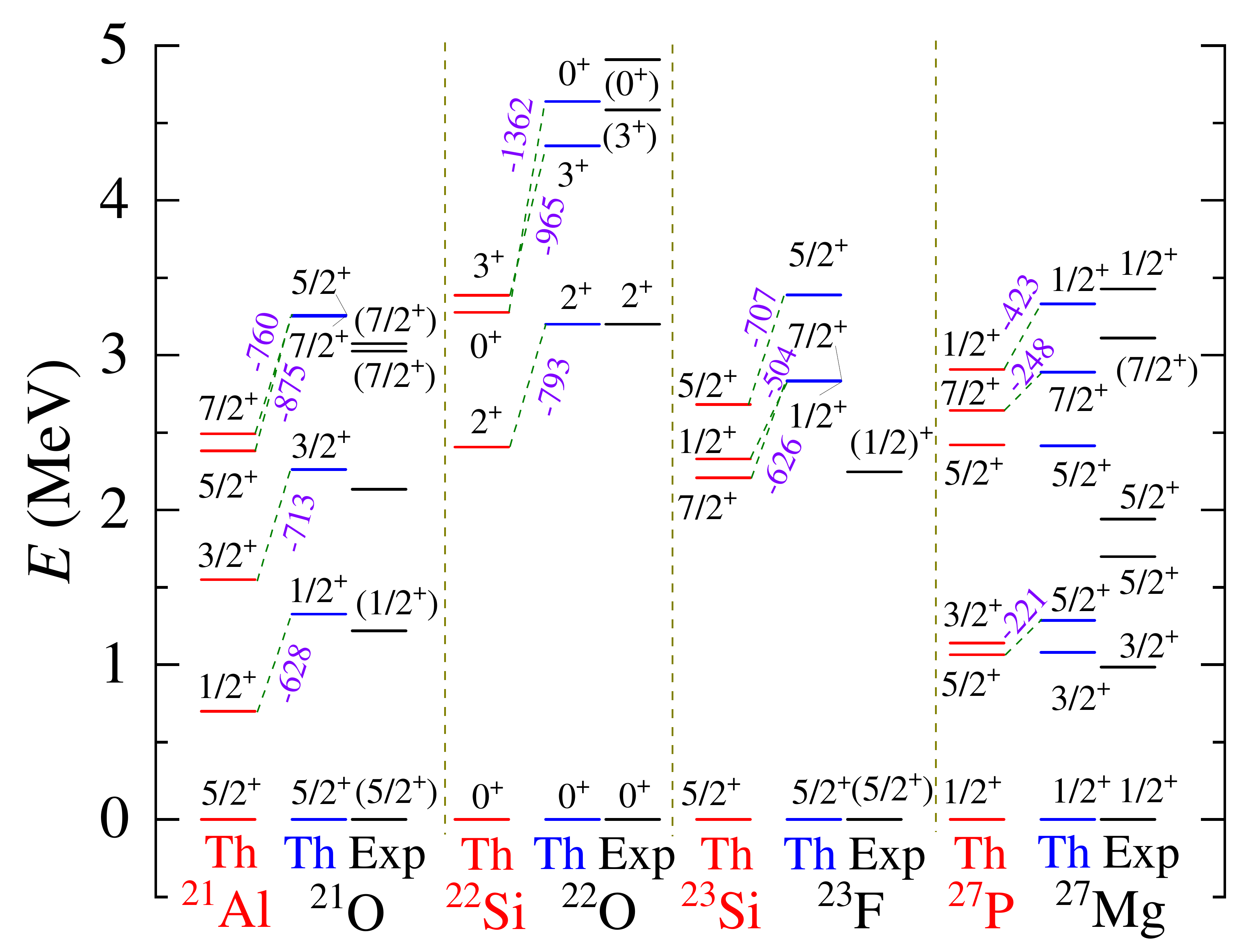}% Here is how to import EPS art
\caption{The spectra (in MeV) of the pairs of mirror nuclei $^{21}$Al/$^{21}$O, $^{22}$Si/$^{22}$O, $^{23}$Si/$^{23}$F, $^{27}$P/$^{27}$Mg. The theoretical spectrum (blue lines) using \textit{ab initio} VS-IMSRG method is compared to the experimental spectrum of neutron-rich nuclei (black lines). Calculations of proton-rich states (red lines) are predictive. Mirror states are connected with green dotted lines and MEDs are marked nearby in purple in keV units.}\label{Predic1}
\end{figure}

Firstly, we calculate the mirror nuclei $^{23}$Al and $^{23}$Ne as an example, due to their rich structure and the fact that several low-lying excited states with large MED exist in their spectra. 
\textit{Ab initio} VS-IMSRG calculations with NN and NN + 3N interactions have then been performed therein. Results are shown in Fig. \ref{23Al}, along with available experimental data \cite{ensdf}. One can see that MED is particularly large in the experimental $1/2^+$ and $3/2^+$ states \cite{ensdf}. The calculation with NN interaction provides a poor description of low-lying states with respect to experimental data. Indeed, even the obtained ground state, which is a $1/2^+$ state in our calculations, is not the experimental one. The calculated energies are, in fact, largely improved by the inclusion of 3N interaction, as a good overall agreement with experiment is obtained for low-lying excited states with NN + 3N interactions. A similar situation also occurs in the other $sd$-shell nuclei studied. Our \textit{ab initio} calculation using the NN + 3N interaction compares well with experimental data.
Hence, \textit{ab initio} VS-IMSRG calculations can be utilized for predictions.
This has been done in Fig. \ref{Predic1} with our model, where the spectra of a few proton-rich nuclei, unknown experimentally, are predicted and compared to their associated mirror spectra. Also, considering that 3N forces are necessary to obtain a satisfactory reproduction of experimental data, we will now only present results calculated with NN + 3N interactions.

%The importance roles of 3N force is clearly shown in the calculations with and without 3N force compared to experimental data. 
%Such that, the calculations only with NN + 3N interaction will be done in the following calculations.

\begin{figure}[!htb]
\includegraphics[width=1.00\columnwidth]{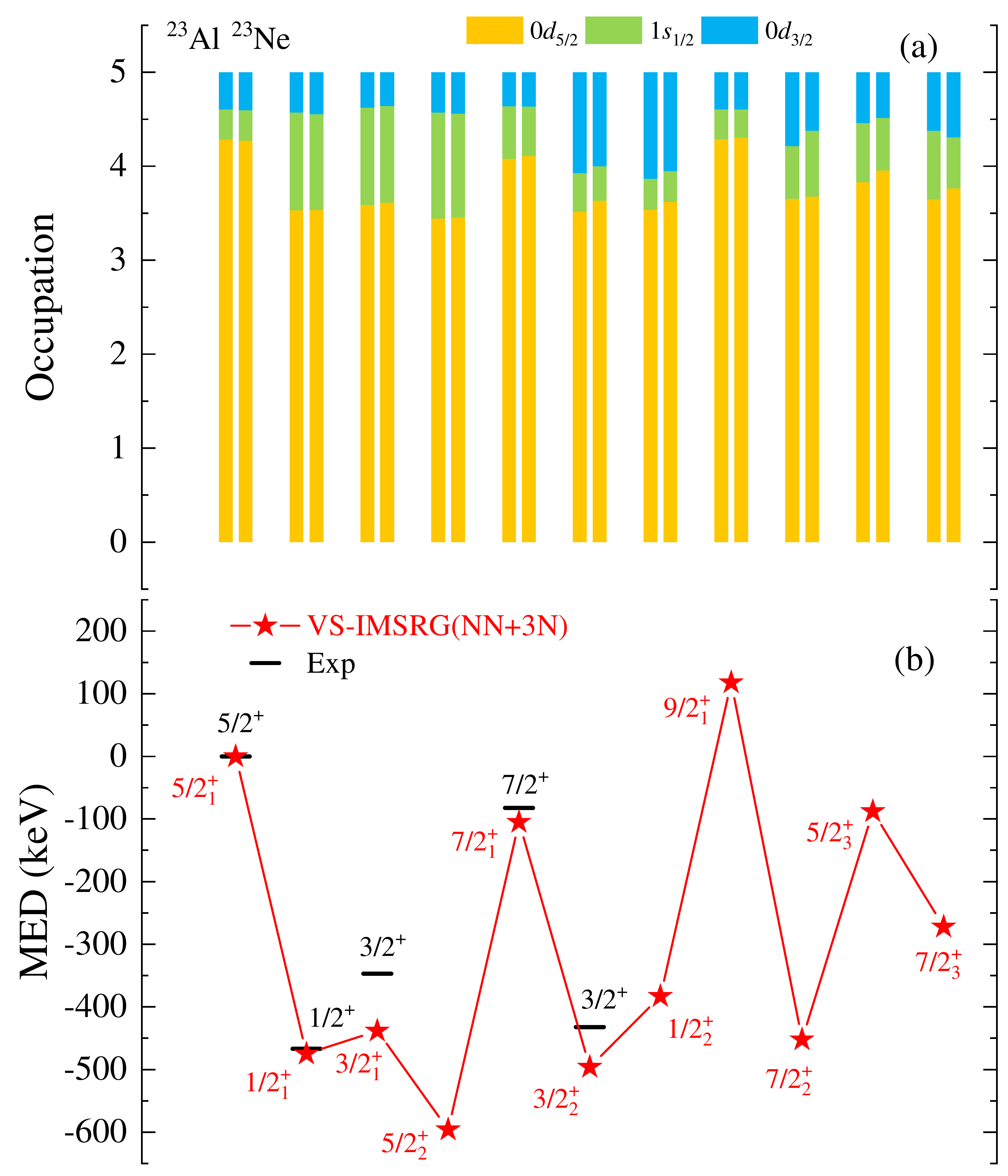}% Here is how to import EPS art
\caption{Calculated average occupations of single-particle valence orbits and mirror energy differences of low-lying states in mirror nuclei $^{23}$Al and $^{23}$Ne using VS-IMSRG based on the 1.8/2.0 (EM) NN + 3N interaction. Shown average occupations correspond to the valence protons and neutrons in $^{23}$Al and $^{23}$Ne, respectively. Experimental data are taken from Ref.\cite{ensdf}.} 
\label{Ocup}
\end{figure}

In Fig. \ref{23Al}, one may notice  that the states bearing the same spin in $^{23}$Al are always lower than those of $^{23}$Ne. For example, the $1/2_1^+$ state of $^{23}$Al is lower than its counterpart in $^{23}$Ne by about 500 keV. 

When a large MED occurs with $s$- or $p$-waves involved, there are two possible reasons for that matter, of external or internal character. If extended single-particle wave functions of weakly- or unbound $s$- or $p$-waves are significantly occupied in the considered many-body states, the large MED is of external nature, as in the TES states \cite{PhysRevC.89.044327,PhysRevLett.125.192503,PhysRevC.90.014307}, neutron skins and halo structures \cite{PhysRevLett.116.212501,PhysRevLett.121.032502}.
The second possibility is related to configuration mixing (see Refs.\cite{PhysRevLett.89.042501,PhysRevLett.89.042502,PhysRevC.102.034302,PhysRevC.104.L061306,LI2022137225}), so that it is of internal nature. In this case, a few nodal states of $s$ or $p$ waves are included in the calculations, so that their mixing generates an extended wave function. Thus, a more extended many-body wave function of dripline nuclei indicates a stronger coupling configuration mixing, i.e.~that internal degrees of freedom are important for that matter. These two external and internal effects are different, but can be intertwined in a complex manner. Configuration mixing involving $s$ and $p$ waves has indeed been proved to describe the exotic properties and MED of dripline nuclei using Gamow shell model (GSM) \cite{0954-3899-36-1-013101,Michel_GSM_book,HU2020135206,PhysRevC.104.024319,PhysRevC.106.L011301}. For example, an extended $1s_{1/2}$ single-particle wave function has been introduced to explain the inversion of ground states in the $^{16}$F and $^{16}$N mirror nuclei \cite{PhysRevC.90.014307}. 
Configuration mixing involving the full $s$ partial wave has also been noticed to play a significant role for that matter,
as GSM calculations have shown \cite{PhysRevC.106.L011301}.

To further analyze the ISB of mirror states, both the average occupations of the single-particle valence orbits and MEDs are presented in the upper and lower panels of  Fig. \ref{Ocup}, respectively, together with available experimental data. 
The MEDs obtained with \textit{ab initio} VS-IMSRG are consistent with the existing  experimental MEDs for the mirror nuclei $^{23}$Al and $^{23}$Ne, as the largest deviation does not exceed 100 keV, as is the case in the $3/2_1^+$ state. Similarly to the previous study, the MEDs of higher lying states, whose experimental values are unknown, are predicted.

The average occupations of the valence proton and neutron shells are presented separately for states in the proton-rich and neutron-rich mirror nuclei in  Fig. \ref{Ocup}.
We can see that the calculated average occupations of valence proton and neutron shells in the mirror states of $^{23}$Al and $^{23}$Ne are very close.
The ground states of $^{23}$Al and $^{23}$Ne are both mainly occupied by the $0d_{5/2}$ orbital. However, the $1/2_1^+$, $3/2^+_1$ and $5/2_2^+$ states bear large $1s_{1/2}$ average occupation, then inducing large MEDs in those states. Indeed, in the mirror nucleus $^{23}$Al, the proton $1s_{1/2}$ state is slightly unbound. Due to the absence of centrifugal barrier in $s$-waves, the proton $1s_{1/2}$ wave function is more extended than that of the neutron $ 1s_{1/2}$ which is deeply bound  \cite{PhysRevC.89.044327}. 
The stronger coupling involving the slightly unbound $\pi 1s_{1/2}$ in $^{23}$Al provides more binding energies than its associated neutron orbital in $^{23}$Ne. This situation results in a negative MED value, i.e.~the excitation energies of mirror states in proton-rich nuclei are lower than those of neutron-rich nuclei due to the strong couplings involving the slightly unbound $s_{1/2}$ waves. For  $7/2_1^+$ states, the occupations are close to those of the  $5/2_1^+$ ground state. The associated MED values are thus small, of about 100 keV.
The large MEDs in the $1/2_1^+$, $3/2^+_1$, and $5/2_2^+$ states show that the TES effect happens in those states.

Interestingly, the $3/2_2^+$ state, which has a large average occupation of $0d_{3/2}$ states, but reduced $1s_{1/2}$ average occupation, exhibits a large MED value, in fact close to that of the $1/2_1^+$ state. The other excited states possessing large $0d_{3/2}$ components also show a similar tendency (see $1/2_2^+$, $7/2_2^+$ and $7/2_3^+$ states in Fig. \ref{Ocup}). This effect occurs because the $0d_{3/2}$ single-particle state is unbound at the proton-rich side and weakly bound at the neutron-rich side of low-lying mirror states. A large average occupation of the $0d_{3/2}$ shell then also accentuates the MED.

Conversely, the $9/2^+$ states provide a MED close to that of the $5/2_1^+$ ground state, as their components are almost only built from $0d_{5/2}$ orbits. The $9/2^+$ MED might then be caused by ISB effects. A similar situation, in fact, commonly occurs in $fp$-shell mirror states \cite{PhysRevLett.87.132502,PhysRevLett.97.132501,PhysRevLett.97.152501,BENTLEY2007497}.
The slight differences in average occupations of valence shells or wave functions of mirror states also contribute to the MED. The situation occurs in the higher mirror states in $^{23}$Al and $^{23}$Ne.

%The proton drip line is closer to the valley of stability, so rich-proton nucleus $^{23}$Al is weakly-bound for well-bound $^{23}$Ne. Hence, ${\pi}(1\textit{s}_{1/2}+0\textit{d}_{3/2})$ especially ${\pi}1\textit{s}_{1/2}(l=0)$ of weakly-bound $^{23}$Al have the resonance and continuum coupling. The continuum effect brings about lower excited energies \cite{ZHANG2022136958}. On the contrary, due to the existence of Coulomb force, the Coulomb barrier which is proportional to angular momentum (\textit{J}) will restrain the reduction of excited states. However \textit{J} equals to zero for \textit{s} wave and Coulomb barrier then is also zero, continuum effect of $1\textit{s}_{1/2}$ orbit dramatically reduce the energy of excited state when the occupation contains large $1\textit{s}_{1/2}$-orbit component \cite{PhysRevC.89.044327}. Although the Coulomb barrier of \textit{d} wave exists, it is small, continuum effect of $0\textit{d}_{3/2}$ orbit thus also can reduce the excited energies.

The indirect effects induced by weakly bound and unbound wave functions on eigenenergies are usually included phenomenologically in the standard shell model by adjusting the matrix elements. For example, in Refs. \cite{PhysRevC.89.044327,PhysRevLett.125.192503}, the matrix elements related to the proton $1s_{1/2}$ orbit are reduced in order to reproduce experimental data.
In order to have physical weakly bound and unbound many-body wave functions, which are extended in coordinate space, it would be preferable to perform shell-model calculation in the Berggren basis, where bound, resonance, and continuum are treated on the same footing, i.e.~in the framework of GSM \cite{PhysRevLett.89.042501,PhysRevLett.89.042502,0954-3899-36-1-013101,Michel_GSM_book,PhysRevC.102.034302,ZHANG2022136958,PhysRevC.104.L061306,LI2022137225,PhysRevC.104.024319}. 
The main drawback of GSM, however, is its computational cost, which is much larger than that of standard shell model \cite{0954-3899-36-1-013101,Michel_GSM_book,HU2020135206,PhysRevC.104.024319}, so that GSM cannot be used in practice for our purposes. 
Consequently, we preferred to consider \textit{ab initio} VS-IMSRG calculation using a large number of HO shells. $N_{\rm max} = 14$ is used in the present calculations, which has been seen to partially describe the extended asymptotes of weakly bound and unbound many-body states, as MEDs could be well described in VS-IMSRG calculations.

\begin{figure}[!htb]
\includegraphics[width=1.00\columnwidth]{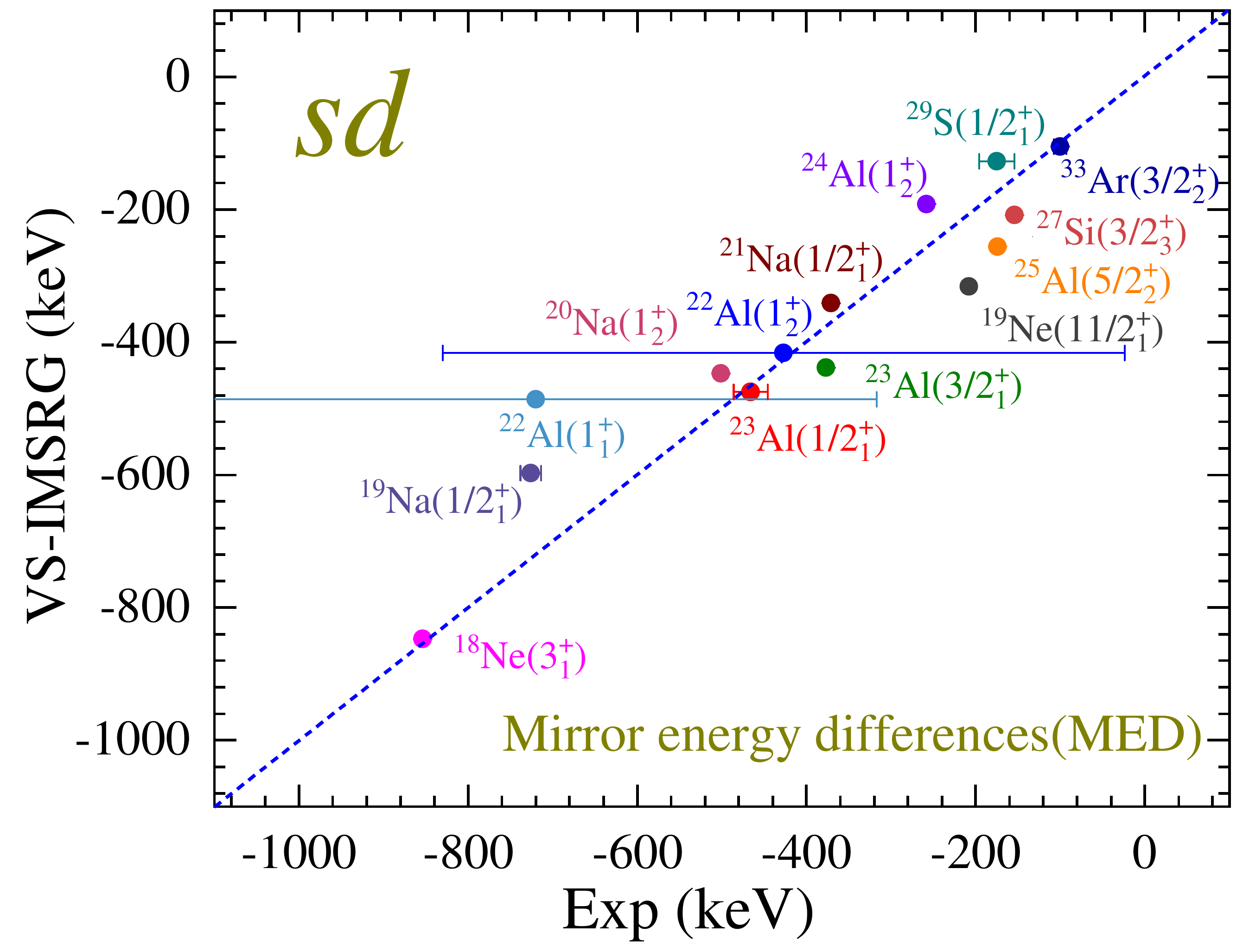}% Here is how to import EPS art
\caption{The calculated MEDs with \textit{ab initio}  VS-IMSRG, labeled with the name of the associated proton-rich nucleus, are compared with experimental data with error bars : $^{18}$Ne($3_1^+$) \cite{PhysRevC.62.055804,PhysRevC.93.044311}, $^{19}$Ne($11/2_1^+$) \cite{PhysRevC.102.045802,PhysRevC.104.025802}, $^{19}$Na($1/2_1^+$) \cite{PhysRevC.67.014308,PhysRevC.93.044311}, $^{20}$Na($1_2^+$) \cite{PhysRevC.86.068801,TILLEY1998249,GLASSMAN2018397},
$^{21}$Na($1/2_1^+$) \cite{PhysRevC.97.065802,b6a73e2789c24f90be2f3ac3011f6002},
$^{22}$Al($1_1^+,1_2^+$) \cite{PhysRevLett.125.192503,PhysRevC.76.034308},
$^{23}$Al($1/2_1^+$) \cite{PhysRevC.64.025802,PhysRevC.104.055806},
$^{23}$Al($3/2_1^+$) \cite{pub.1101152511,doi:10.1063/1.3087047},
$^{24}$Al($1_2^+$) \cite{Ichikawa2011ProtonrichNS,PhysRevC.89.014617},
$^{25}$Al($5/2_2^+$) \cite{PhysRevC.92.031302,PhysRevC.104.055806},
$^{27}$Si($3/2_3^+$) \cite{PhysRevC.94.065806,PhysRevC.84.035802},
$^{29}$S($1/2_1^+$) \cite{PhysRevC.81.067303,PhysRevC.94.064305},
$^{33}$Ar($3/2_2^+$) \cite{PhysRevLett.92.172502,PhysRevC.30.1442}. }
\label{MED}
\end{figure}

To further show the predicting power of our \textit{ab initio} VS-IMSRG calculations, we have performed calculations of MEDs in a large set of $sd$-shell nuclei. The pairs of states with the largest experimental MED are presented in Fig. \ref{MED}. 
The distance between points and the diagonal line in Fig.~\ref{MED} is the criterion for determining whether \textit{ab initio} VS-IMSRG calculations are in accordance with experiment. In fact, all points are situated near the diagonal line in Fig. \ref{MED}. The largest deviations occur in $^{19}$Na$(1/2_1^+)$ and $^{22}$Al$(1_1^+)$ and do not exceed 150 keV. Hence, \textit{ab initio} VS-IMSRG method can provide a good description of the properties of $sd$-shell nuclei, so that it can be now applied  for predictions related to many-body states inaccessible experimentally.

In Fig. \ref{Predic1}, we give the predictions of unmeasured states at proton dripline, namely $^{21}$Al, $^{22}$Si, $^{23}$Si and $^{27}$P, along with the experimental and theoretical results related to their mirror neutron-rich nuclei for comparison. 
From Fig. \ref{Predic1}, we can see that the \textit{ab initio} VS-IMSRG calculations of the neutron-rich nuclei $^{21}$O, $^{22}$O, $^{23}$F and $^{27}$Mg are in good agreement with the experimental data. Therefore, one can assume that the prediction for the spectra of the aforementioned proton-rich nuclei issued from \textit{ab initio} VS-IMSRG calculations is reliable. The predicted MEDs of most associated mirror states are about a few hundred keV units and are negative. Notably, a large MED ($-1362$ keV) is present in the $0^+_2$ excited states of $^{22}$Si and $^{22}$O. MEDs significantly increase when proton-rich nuclei approach the proton dripline, as the larger MEDs occur when the average occupation of weakly bound or unbound \textit{s} orbitals increase (see the discussion related to Fig. \ref{Ocup} for details). We also provided predictions for possible excited states in $^{21}$Al, $^{22}$Si, $^{23}$Si and $^{27}$P proton-rich nuclei for experiment. We hope that our calculations will be useful in future experiments searching for new nuclear excited states in light nuclei at drip-lines.

\begin{figure}[!htb]
\includegraphics[width=1.00\columnwidth]{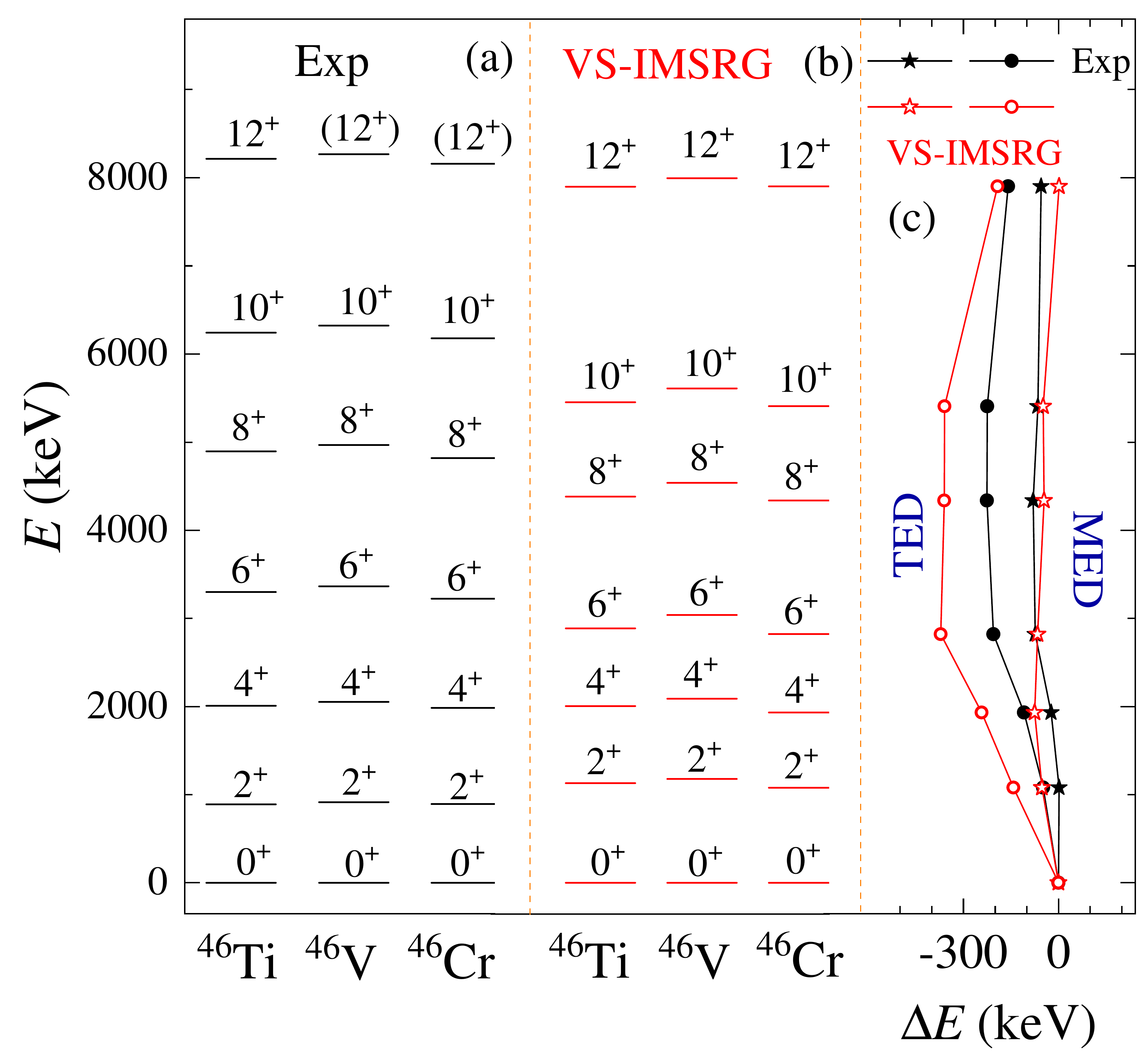}% Here is how to import EPS art
\caption{Yrast states with $T=1$ in $^{46}$Ti, $^{46}$V and $^{46}$Cr. (a) provides their excitation energies from experimental data, taken from Ref.~\cite{PhysRevLett.87.132502}. (b) shows the calculated excitation energies using \textit{ab initio} VS-IMSRG with the 1.8/2.0 (EM) NN + 3N interaction. (c) shows MED and TED (mirror and triplet energy difference) values as a function of the angular momentum $J$ (the shown angular momenta correspond to the calculated angular momenta in the spectrum of $^{46}$Cr). The black full stars (red open stars) denote the experimental (\textit{ab initio} VS-IMSRG) MED values, while black full dots (red open dots) refer to the experimental (\textit{ab initio} VS-IMSRG) TED values.}  {}
\label{T=1_46_MED}
\end{figure}

As can be observed from experimental data, MEDs vary with total spin $J$ in a given spectrum \cite{PhysRevLett.92.132502,PhysRevLett.97.132501,PhysRevLett.97.152501,BENTLEY2007497,PhysRevLett.109.092504,PhysRevLett.117.082502,YAJZEY2021136757,PhysRev.116.465}.
Systematic studies of the behavior of MED as a function of angular momentum have been done in the $f_{7/2}$ shell up to rather a high spin.
Theoretical standard shell-model calculations, including a range of electromagnetic effects  as well as a schematic isospin-non-conserving interaction, have been employed to investigate the MED in $pf$-shell nuclei \cite{PhysRevC.92.024310,BENTLEY2007497,PhysRev.116.465}. 
To test our \textit{ab initio} VS-IMSRG calculations of mirror and triplet energy differences (MED and
TED) involving many-body states whose angular momentum reaches $J \sim 10$, the $^{46}$Ti, $^{46}$V and $^{46}$Cr nuclei have been considered as a testing ground (see Fig. \ref{T=1_46_MED}).
MED is usually small in $fp$-shell nuclei when compared to the nuclear states of the $sd$-shell nuclei. Quantitatively, the MED values are typically smaller than 100 keV. The Coulomb and isospin-non-conserving forces are responsible for these values, whereas the effects arising from the extension of wave functions, prominent in weakly bound and unbound state bearing sizable $s$-wave average occupation, can be neglected.

The MED and TED have been investigated as functions  of the angular momentum $J$. 
The MED and TED are equal to $\textit{E}_{ex}(T_z=-1,J^{\pi}) - \textit{E}_{ex}(T_z=1,J^{\pi})$ and $\textit{E}_{ex}(T_z=-1,J^{\pi}) + \textit{E}_{ex}(T_z=1,J^{\pi}) - 2\textit{E}_{ex}(T_z=0,J^{\pi})$, respectively. $E_{ex}$ denotes the excited energy, so that the above definitions allow to remove dependence on the ground-state energy.
The results with \textit{ab initio} VS-IMSRG based on NN + 3N interaction are presented in Fig. \ref{T=1_46_MED}. 
From Fig. \ref{T=1_46_MED}(a) and Fig. \ref{T=1_46_MED}(b), we can see that the calculated spectra are very satisfactory when compared to experimental energies. 
In order to investigate the ISB, we plot the MED and TED values as functions of angular momentum $J$. The results are shown in Fig. \ref{T=1_46_MED}(c) and compared to experimental values.
Both the experimental and theoretical values show that the MED and TED are of the order of several tens and hundreds of keV, respectively.
The MED values show satisfactory agreement between \textit{ab initio} VS-IMSRG calculations and experimental data, although there is a small discrepancy for $J=2,4$ many-body states. These results show better agreement with experimental data compared to standard shell model calculations.
For TED, our calculations are always larger about 100 keV than experimental data in magnitude. Note that a similar situation occurs in realistic shell model calculations \cite{PhysRevC.96.024323}.
However, the variation of experimental TED with spin is reproduced in \textit{ab initio} VS-IMSRG calculations. Standard shell model calculations  provided a good agreement with experimental data for the TED values up to $J=12$ \cite{BENTLEY2007497}.
Further studies are thus needed in standard shell model and in our \textit{ab initio} calculations.

\subsection{Isobaric Multiplet Mass Equation}

Another ISB signature is the IMME. Wigner \cite{Wigner_IMME}, Weinberg and Treiman \cite{PhysRev.116.465} indeed noted that the masses of isospin multiplets with the same mass number $A$, total isospin $T$ and spin-parity $J^{\pi}$ but different isospin third component $T_z$ satisfy the relationship called IMME, written as
\begin{equation}
\operatorname{ME}\left(A, T, T_{z}\right)=a+b T_{z}+c T_{z}^{2}, \label{IMME}
\end{equation}
where ME is mass excess, and $a$, $b$, and $c$ are coefficients. 
%The equation is obtained by the Wigner-Eckart theorem and in the first-order perturbation theory. 
$T=3/2$ isospin quartets and larger isospin multiplets enable us to test the IMME equation \cite{PhysRevLett.109.102501, LAM2013680, 2014PhRvL.113h2501G, PhysRevC.99.014319, PhysRevC.80.051302, PhysRevC.84.031301}. The present paper mainly focuses on $T=1/2$ isospin doublets and $T=1$ isospin triplets. When the isospin $T= 1/2$, the $a$ and $b$ coefficients can be calculated from the following formulas :
\begin{equation}
\begin{aligned}
a=\operatorname{ME}\left(T_z=1/2\right)+\operatorname{ME}\left(T_z=-1/2\right),\\
b=\operatorname{ME}\left(T_z=1/2\right)-\operatorname{ME}\left(T_z=-1/2\right).
 \label{T1/2}
 \end{aligned}
\end{equation}
For the isospin $T=1$, the $a$, $b$ and $c$ coefficients can be obtained from the equalities :
\begin{equation}
\begin{aligned}
a=&\operatorname{ME}\left(T_z=0\right),\\
b=&\left(\operatorname{ME}\left(T_z=1\right)-\operatorname{ME}\left(T_z=-1\right))\right/2,\\
c=&\left(\operatorname{ME}\left(T_z=1\right)+\operatorname{ME}\left(T_z=-1\right))\right/2-\operatorname{ME}\left(T_z=0\right).
 \label{T1}
 \end{aligned}
\end{equation}

To compare with Coulomb contribution, the Coulomb energy with classical and semi-classical approach has been also calculated. The classical approach treats a nucleus as a uniformly charged sphere \cite{RevModPhys.8.82} of radius $R=r_0A^{1/3}$, so that the Coulomb energy reads : 
\begin{equation}
\begin{aligned}
E_{\text{Coul}}^{\text{classical}}&=\frac{3 e^{2}}{5 R} Z(Z-1),\\
&=\frac{3 e^{2}}{5 r_{0} A^{1/3}}\left[\frac{A(A-2)}{4}+(1-A) T_{z}+T_{z}^{2}\right].
 \label{classicalCou}
 \end{aligned}
\end{equation}
The Coulomb energy from semi-classical approach is obtained by adding the direct term, equal to $(3/5R)e^2Z^2$, and the exchange term evaluated using plane waves for proton wave functions \cite{doi:10.1146/annurev.ns.19.120169.002351,Shlomo_1978} :
\begin{equation}
\begin{aligned}
E_{\text{Coul}}^{\text{semi-classical}}=&\left\{0.6 Z^{2}-0.46 Z^{4 / 3}-0.15\left[1-(-1)^{Z}\right]\right\}\\ &\times \frac{e^{2}}{r_{0} A^{1/3}},
 \label{semiclassicalCou}
 \end{aligned}
\end{equation}
in which both the exchange effects and Coulomb pairing energies are included. The value $r_0=1.25$ fm is used in the present work. The $b$ coefficient can be calculated from Coulomb contributions and is equal to $b=E_{\text{Coul}}(T_z =1/2)-E_{\text{Coul}}(T_z=-1/2)+(M_n-M_p)$ for $T=1/2$ and $b=(E_{\text{Coul}}(T_z=1)-E_{\text{Coul}}(T_z=-1))/2+(M_n-M_p)$ for $T=1$, where $M_n$ is the neutron mass and $M_p$ represents the proton mass, the neutron-proton mass difference being equal to 782 keV.

\begin{figure}[!htb]
\includegraphics[width=1.00\columnwidth]{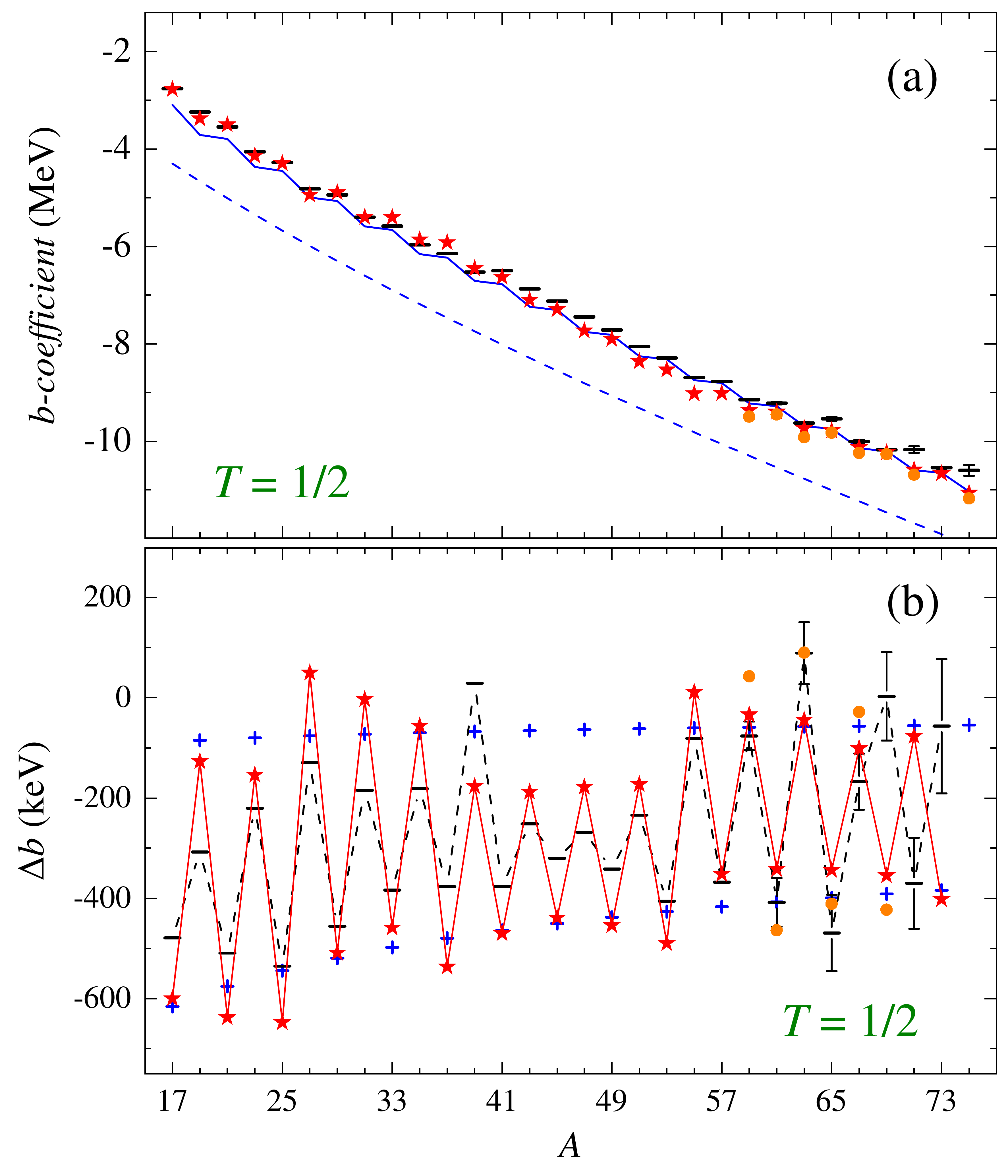}% Here is how to import EPS art
\caption{Calculations of the $b$ coefficient of IMME in MeV (upper panel)  and of the differences of $b$ values between nuclei $A$ and $A+2$ in keV, denoted as $\Delta b$ (lower panel), with $T=1/2$ in nuclei bearing $A=17$ to $A=75$ nucleons. The results of \textit{ab initio} VS-IMSRG calculations using a $^{16}$O or $^{40}$Ca inner core based on the 1.8/2.0 (EM) NN + 3N interaction are labelled with red stars. The orange points denote calculations using $^{56}$Ni as an inner core. Theoretical results are compared with experimental data collected from AME2020 \cite{Wang_2021}, which is indicated by black lines. The blue dotted line in the upper panel arises from the classical Coulomb formula of Eq. (\ref{classicalCou}). The blue full line in the upper panel and unconnected blue plus signs in the lower panel are calculated using the semi-classical Coulomb formula of Eq. (\ref{semiclassicalCou}).}
\label{T0.5}
\end{figure}   

Figure \ref{T0.5}(a) shows the results of the IMME $b$ coefficient calculated from $T=1/2$ isospin doublets  for mass number from 17 to 75 using \textit{ab initio} VS-IMSRG, along with experimental data and results of Coulomb contribution using Eqs. (\ref{classicalCou}) and (\ref{semiclassicalCou}).
We also performed VS-IMSRG calculations of the upper $fp$-shell nuclei within the $0f_{5/2},1p_{3/2},1p_{1/2},0g_{9/2}$ valence space, hence with both valence protons and valence neutrons above a $^{56}$Ni core.
Overall, our VS-IMSRG results satisfy AME2020 values, where the largest differences do not exceed 300 keV for the $b$ coefficient. 
Similar results are also obtained for the upper $fp$-shell nuclei with two different reference cores, i.e.~$^{40}$Ca and $^{56}$Ni (see Fig. \ref{T1}). 
For Coulomb energies, the results from the semi-classical approach (See Eq. (\ref{semiclassicalCou})) provide good agreement with experimental data. However, the $b$ coefficient issued from the classical approach is more attractive compared to its experimental value. Moreover, the staggering pattern could also be described with our VS-IMSRG method and semi-classical approach of Coulomb energy but not with the classical approach of Coulomb energy.  

To see the staggering pattern more clearly, the differences of $b$ coefficient in nuclei $A$ and $A+2$ can be computed using 
\begin{equation}
\Delta b(A, T)=b(A+2, T)-b(A, T).
 \label{chab}
\end{equation}
The calculated $\Delta b$ values using VS-IMSRG and semi-classical approach of Coulomb energy are presented in Fig.~\ref{T0.5}(b), along with experimental data. Our VS-IMSRG calculations clearly reproduce the experimentally observed odd-even staggering pattern. 
Moreover, its drastic damping in $\textit{f}_{7/2}$-shell nuclei has also been obtained in our calculations, which is assumed to originate from shell effects \cite{PhysRevC.103.024316}.
Finally, it is worth noting that the calculations starting from $\textit{A}=69$ do not support the staggering phase which is observed experimentally. A similar situation also occurs in shell-model calculations \cite{PhysRevLett.110.172505}, where it is suggested that the mass of $^{69}$Br was, in fact, measured for an isomeric state, and not for its ground state \cite{PhysRevLett.110.172505}. We suppose that the discrepancy of $\Delta b$ from $A=69$ to 73 was caused by the large differences of  $b$-coefficient of $A=71$ and 75, in which the mass of $^{71}$Kr and $^{75}$Sr are taken from AME2020 extrapolation \cite{Wang_2021}. Further mass measurements are thus needed to clarify the situation.

%And the zigzag of IMME's $b$ coefficient appears at $T=1/2$ and disappears at $T=1$ from Fig. \ref{T0.5} and Fig. \ref{T1}, as the conclusion of Ref. \cite{PhysRevC.103.024316} gives that the staggering effect for $b$ coefficient is visible on half-integer $T$ and is negligible on integer $T$. Finally, it's worth noting that the calculations starting from $\textit{A}=69$ don't support the variational staggering phase of the experiment (also seen in Ref \cite{PhysRevLett.110.172505}). The situation suggests that further precious measured masses should be approved and  theoretical calculation need to further improvement.

\begin{figure}[!htb]
\includegraphics[width=1.00\columnwidth]{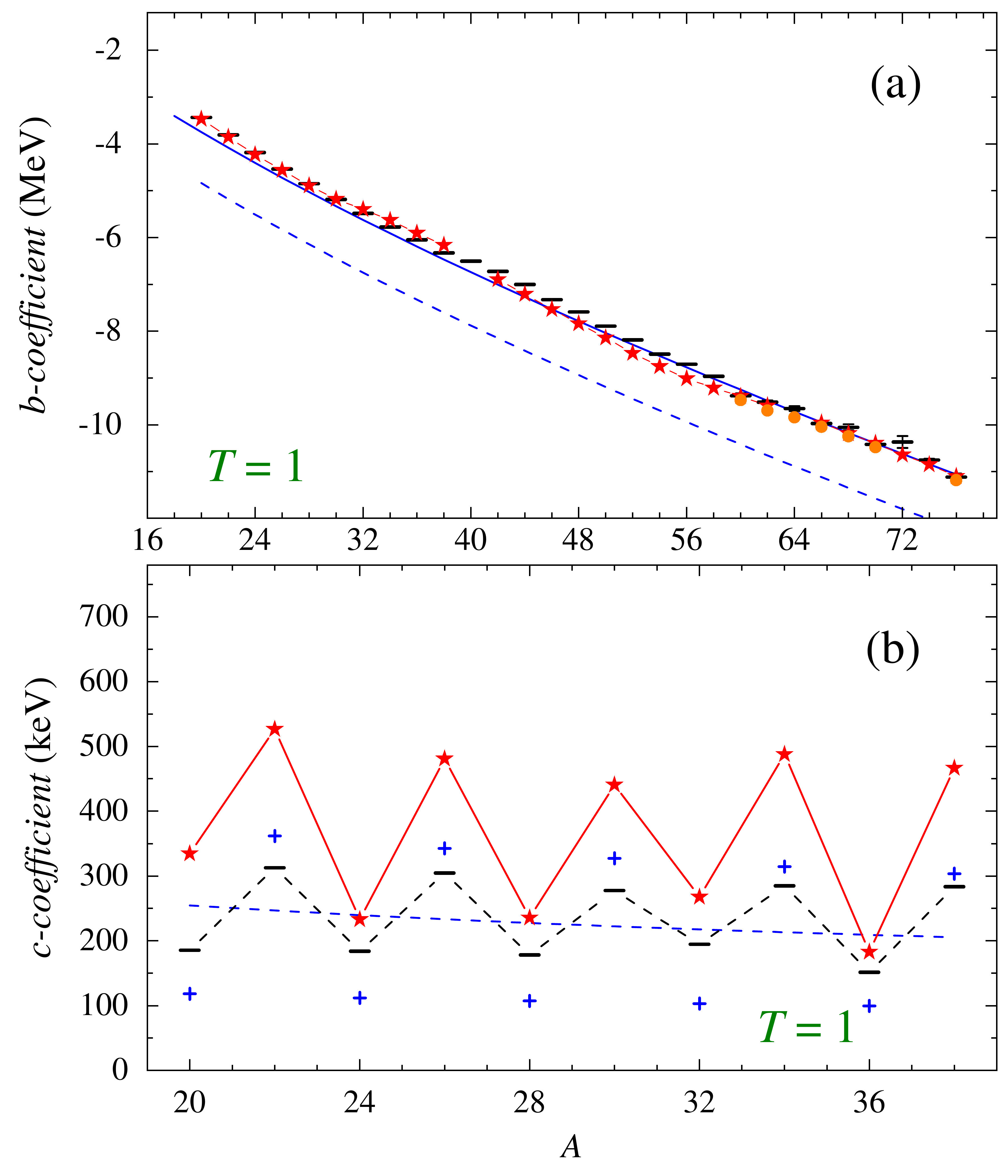}% Here is how to import EPS art
\caption{Similar to Fig. \ref{T0.5}, but for the IMME $b$ coefficient (in MeV unit) determined with the $T=1$ nuclei from $A=20$ to $A=76$ (upper panel) and IMME $c$ coefficient (in keV unit) calculated with nuclei with $A$ varying from 20 to 38 (lower panel). 
%The Coulomb energy in the upper panel is obtained using Eqs. (\ref{classicalCou}) and (\ref{semiclassicalCou}). Experimental data in the lower panel are taken from Ref.~\cite{ensdf}. 
}{}
\label{T1_b}
\end{figure}

Similar calculations to determine the $b$ coefficient in $T=1$ isospin triplets for $A = 20-76$ have also been performed. Results are shown in Fig. \ref{T1_b}(a) and compared with experimental data and calculated Coulomb energies. The calculated $b$ coefficient from our VS-IMSRG and semi-classical approach of Coulomb energy are in good agreement with experimental data. The classical approach of Coulomb energy gives overbound $b$ values, however, similarly to the $T=1/2$ case.  

The $c$ coefficient of $T=1$ isospin triplets has also been considered. Due to the huge computational cost of full diagonalization of $pf$-shell nuclei, only the $c$ coefficients of $sd$-shell nuclei have been computed (see Fig.  \ref{T1_b}(b)). Our VS-IMSRG calculations provide $c$ coefficients which are systematically larger than experimental data, especially for odd values of $A/2$. 
The same situation is also obtained in Ref.  \cite{PhysRevC.104.014324}. 
However, the oscillatory pattern of the $c$ coefficient in $T=1$ isospin triplets is well reproduced in our VS-IMSRG calculations, contrary to the results obtained with the macroscopic-microscopic approach \cite{PhysRevC.103.024316}.
Larger $c$ coefficients are also obtained in shell-model calculations compared with experimental data, where isospin-non-conserving and Coulomb effects are considered within many-body perturbation theory (MBPT) \cite{PhysRevC.96.024323}.  
Good agreement with experimental data is nevertheless obtained when including the bare Coulomb and isospin-non-conserving force without MBPT renormalization \cite{PhysRevC.96.024323}. As for the IMME $b$ coefficient, more advanced theoretical calculations and nucleon-nucleon interactions might be necessary to have a more precise determination of the $c$ coefficient, as discussed in Ref. \cite{PhysRevC.96.024323} for example.

\section{Summary}

Isospin-symmetry breaking effects in mirror energy difference and nuclear mass have been investigated using the \textit{ab initio} valence-space in-medium similarity renormalization group approach based on the chiral interactions.
Charge-symmetry breaking and charge-independent breaking, which contribute to the isospin-non-conserving part of nuclear forces, are included in the nuclear potential defining the effective Hamiltonian of \textit{ab initio} calculations. The Coulomb force clearly has to be included in these calculations.

Firstly, the \textit{ab initio} VS-IMSRG with NN and NN + 3N interactions are performed for the $sd$-shell nuclei, in which the $^{23}$Al and $^{23}$Ne are taken as an example to study the mirror energy difference, with results compared to experimental data.
From the VS-IMSRG calculations of low-lying states in $^{23}$Al and $^{23}$Ne, it has been shown  that the three-body force plays an essential role in order to obtain well reproduced excited states. 
The MEDs have been investigated and discussed  thoroughly in the mirror nuclei $^{23}$Al/$^{23}$Ne, by analyzing the  occupation of weakly bound and unbound single-particle states for generating TES, especially the $1\textit{s}_{1/2}$ orbit. 
A comparison of MEDs bearing large values in experiments with theoretical VS-IMSRG calculations have been done  for $sd$-shell nuclei, and predictions about proton-rich nuclei which are inaccessible experimentally have been made. 
The $J$ dependence in MED and TED for $pf$-shell $A=46$ nuclei has been analyzed as well and has been shown to be small in the spectrum of a fixed pair of mirror nuclei.
Finally, we also calculated the IMME coefficients in $T=1/2$ doublets with $A$ from 17 to 75 and in $T=1$ triplets with $A$ from 20 to 76. 
The agreement with experimental data is quite satisfactory for the $b$ coefficients in $T=1/2$ and $T=1$ nuclei. 
The experimental  $c$ coefficient is qualitatively well described by our \textit{ab initio} VS-IMSRG calculations, as the experimentally observed odd-even staggering is present at theoretical level, while the calculated  absolute values differ by at most about 200 keV from experimental data. 
As a whole, the isospin-symmetry breaking phenomena could be well described with \textit{ab initio} VS-IMSRG calculations. However, the observed small discrepancies still need to be improved by developing more accurate nuclear forces and more advanced many-body techniques.

\textit{Acknowledgments.}~
 This work has been supported by the National Natural Science Foundation of China under Grant Nos.  12205340, 12175281, 11975282, 11921006, 11835001 and 12035001;  the Gansu Natural Science Foundation under Grant No. 22JR5RA123;  the Strategic Priority Research Program of Chinese Academy of Sciences under Grant No. XDB34000000; the Key Research Program of the Chinese Academy of Sciences under Grant No. XDPB15; the State Key Laboratory of Nuclear Physics and Technology, Peking University under Grant No. NPT2020KFY13. This research was made possible by using the computing resources of Gansu Advanced Computing Center. 

\bibliography{Ref}
\end{document}